\documentclass[10pt]{article}
\usepackage{amsmath, amsfonts, amssymb, graphicx, color, float, physics}
\usepackage[colorlinks=true,citecolor=blue,linkcolor=blue,urlcolor=blue]{hyperref}
\usepackage[numbers,sort&compress]{natbib}
\usepackage{geometry}
\begin{document}

\title{\bf{Measurement-Induced Local Dephasing Generates Symmetrically Located Entangled Sites in a Fermionic Tight-Binding Lattice}}

\author{
Aitijhya Saha$^{1,2}$, Srijani Das$^{1,3}$, and Debraj Rakshit$^{1}$
}

\date{
\small
$^{1}$Harish-Chandra Research Institute, A CI of Homi Bhabha National Institute, Allahabad 211019, India\\
$^{2}$Department of Theoretical Physics, Tata Institute of Fundamental Research, Mumbai 400005, India\\
$^{3}$Physics and Applied Mathematics Unit, Indian Statistical Institute, Kolkata 700108, India
}


\maketitle

\subsection*{Abstract}

We investigate an odd-sized fermionic open tight-binding chain subjected to stochastic projective measurements at its central site, effectively inducing localized dephasing. Focusing initially on the single-particle regime, we demonstrate that when the system is prepared in an even-parity state, the dynamics under central-site dephasing drive it toward a nontrivial steady state, which we characterize through both analytical and numerical approaches. Remarkably, this steady state exhibits long-range quantum correlations in the form of symmetrically positioned, pairwise entangled sites across the chain. We further show that the degree of pairwise mode entanglement can be significantly enhanced by increasing the particle number, provided the system is initialized within a specific symmetry sector associated with an underlying strong symmetry operator. Our results identify a minimal measurement-induced route for generating symmetry-selected long-range pairwise mode entanglement, with possible implications for quantum communication and distributed quantum information processing. 

\subsection*{1. Introduction}
The shareability of quantum correlations is fundamentally constrained. In particular, the monogamy of entanglement limits the extent to which bipartite quantum correlations can be distributed across a multipartite system \cite{Tomamichel,Dhar17}. For a pure state of $N$ qubits $\rho_{A_1A_2 \dots A_N}$, the generalized Coffman--Kundu--Wootters (CKW) inequality \cite{Coffman} reads $\mathcal{C}^2_{A_1:(A_2 \dots A_N)} \geq \sum_{j=2}^N \mathcal{C}^2_{A_1 A_j}$, where $\mathcal{C}_{A_1:(A_2 \dots A_N)}$ denotes the concurrence between qubit $A_1$ and the rest of the system, and $\mathcal{C}_{A_1 A_j}$ quantifies pairwise concurrence. This relation encapsulates the trade-off in entanglement distribution: increased bipartite entanglement with one subsystem necessarily suppresses correlations with others.

Such constraints strongly influence entanglement structure in quantum many-body systems. Long-range entanglement is, however, a central resource for quantum information processing. Protocols such as teleportation, dense coding, distributed quantum computation, and quantum communication require bipartite or multipartite entanglement shared over spatially separated nodes.  In paradigmatic models such as quantum spin chains and Hubbard systems, pairwise entanglement in typical low-energy states decays rapidly with distance and becomes negligible beyond nearest neighbors \cite{Osterloh02, Sadhukhan16}. This presents a major bottleneck for implementing entanglement-assisted protocols between distant nodes of a quantum network. To address this limitation, several proposals have been put forward to engineer entanglement between arbitrary pairs of lattice sites, particularly aimed at generating long-range entanglement. This includes, e.g., entanglement swapping, repeaters, or engineered local Hamiltonians with spatial modulation \cite{Zukowski93,Samara,Briegel98,Santra,Ruihong19,Poop05,Sadhukhan17,Amaro,Ramirez15,Venuti06, Dhar16}. 

Interestingly, structured long-range entangled steady states can also be generated by subjecting the system to a specifically designed noisy environment, e.g., via spatially localized dissipative pairing interaction or via local loss-gain mechanisms \cite{Pocklington22, Dutta20, Dutta24}. This is counterintuitive, as entanglement, in general, is known to be fragile under open dynamics, posing a major challenge for controlled quantum-information processing.\cite{Palma96,Hartmann,Buchleitner09,Isar,Nielsen10,Zurek06,Aolita}. However, many recent works have demonstrated experimentally realizable useful quantum resource generation and preparation of certain unique states via clever engineering of the coupling with the environment \cite{Plenio02,Kraus08,Krauter11,Poyatos96,Diehl08,Muller12,Rai10,Schirmer10,Diehl11,Verstraete09,Dutta20, Ray23, Dutta21}. 

These developments motivate a natural question: whether long-range entanglement can be generated and stabilized dynamically using only minimal local operations. We address this question in an odd fermionic tight-binding chain, showing that stochastic projective measurements at a single central site, equivalently local dephasing, can generate steady states containing pairwise entanglement between mirror-related distant sites. The key ingredient is symmetry -- a foundational organizing principle in quantum systems, fundamentally shaping their structure, dynamics, and emergent phenomena ~\cite{Elze09,Zee07,Cariglia14,Daily12,Thingna16,Thingna20}. Symmetry protection can, in principle, preserve information about the initial state and has implications for robust quantum memory and state engineering; thus, understanding its role in open dynamics is fast becoming a central topic of research in recent times~\cite{Breuer07,Muller11, Rivas12,Bairey20,Baumgartner08, Buca12, Albert14, Manzano14, Albert16, Dutta20, Dutta24}. On the other hand, measurement-induced dynamics has emerged as a distinct paradigm for controlling quantum many-body systems \cite{Li18, Skinner19, Chan19, Feng23,Dubey23}. In particular, stochastic sequences of projective measurements give rise to effective non-unitary evolution, often describable as measurement-induced dephasing, while preserving underlying conservation laws \cite{Facchi08, Gherardini15, PascazioReview}.

The stochastic projective measurements at the central site induce local dephasing without introducing particle loss or gain, thereby preserving number conservation. We show that the resulting dynamics is constrained by a strong symmetry operator, which partitions the Hilbert space into invariant sectors characterized by conserved charges. We identify classes of symmetry-resolved initial states that evolve toward well-defined steady states under this central-site measurement protocol. In the single-particle regime, the relevant symmetry reduces to parity, and we demonstrate that even-parity initial states lead to a unique steady state exhibiting pairwise mode entanglement between sites symmetrically located about the center. In the many-particle case, the strong symmetry becomes distinct from parity and plays a decisive role: Long-range entangled pairs occur specifically in the irreducible sector with the maximal positive charge. We provide explicit analytical expressions for the steady states, along with closed-form results for pairwise correlations and mode entanglement, supported by numerical simulations.

The remainder of the paper is organized as follows. In Sec.~2, we introduce the stochastic non-selective measurement protocol considered in this work. In Sec.~3, we formulate the protocol within the framework of open quantum systems and establish its equivalence to a realization involving Gaussian white noise. In Sec.~4, we discuss the concept of strong symmetry and its consequences for the system dynamics. In Sec.~5, we establish the uniqueness of the steady state in the single-particle sector and characterize the steady state explicitly. In Sec.~6, we investigate the emergence of pairwise mode entanglement between lattice sites in the single-particle steady state. In Sec.~7, we extend our analysis to the multi-particle sector. Finally, Sec.~8 summarizes our main results and provides an outlook for future directions.

\subsection*{2. Stochastic Non-Selective Measurement Protocol}

We consider a single particle evolving on a one-dimensional tight-binding lattice with Hamiltonian $\hat{H}$. A detector is placed at a fixed lattice site, and repeated projective measurements are performed to probe the presence of the particle at that site.

The measurement times are stochastic. Specifically, the waiting times between successive measurements, $\{\tau_1, \tau_2, \dots\}$, are independent and identically distributed random variables drawn from the exponential distribution
\begin{equation}
p(\tau) = \gamma e^{-\gamma \tau},
\label{eq:one}
\end{equation}
where $\gamma$ denotes the measurement rate. The total observation time is fixed to $t$. For a given realization consisting of $m$ measurements, the first $m-1$ intervals are sampled from $p(\tau)$, while the final interval is set by $\tau_m = t - \sum_{i=1}^{m-1} \tau_i$. The system is initialized in a pure state $\hat{\sigma}(0) = |\Psi(0)\rangle\langle\Psi(0)|$.
Between successive measurements, the system evolves unitarily according to
$\hat{U}(\tau) = e^{-i \hat{H} \tau}$.

At each measurement time, a projective measurement is performed at the detector site. Let $\hat{\Pi}_d$ denote the projection operator onto the detector site, and $\hat{\Pi}_{\bar{d}} = \mathbb{I} - \hat{\Pi}_d$ the complementary projector. The measurement thus yields two possible outcomes: detection ($d$) or non-detection ($\bar{d}$). The post-measurement state, conditioned on outcome $x \in \{d, \bar{d}\}$, is given by
\begin{equation}
\hat{\sigma} \rightarrow \frac{\hat{\Pi}_x \hat{\sigma} \hat{\Pi}_x}{\mathrm{Tr}(\hat{\Pi}_x \hat{\sigma})}.
\end{equation}

A single realization of the protocol is specified by (i) a sequence of waiting times $\{\tau_1,\dots,\tau_m\}$ and (ii) a sequence of measurement outcomes $\boldsymbol{x}=\{x_1, x_2,\dots,x_m\}$, with $x_j \in \{d,\bar{d}\}$. The corresponding (unnormalized) density matrix after $m$ measurements is
\begin{equation}
\tilde{\sigma}^{(\boldsymbol{x})}(t) =
\hat{\Pi}_{x_m} \hat{U}(\tau_m)
\cdots
\hat{\Pi}_{x_2} \hat{U}(\tau_2)
\hat{\Pi}_{x_1} \hat{U}(\tau_1)
\hat{\sigma}(0)
\hat{U}^\dagger(\tau_1) \hat{\Pi}_{x_1}
\hat{U}^\dagger(\tau_2) \hat{\Pi}_{x_2}
\cdots
\hat{U}^\dagger(\tau_m) \hat{\Pi}_{x_m}.
\label{eq:three}
\end{equation}

Expectation values of observables require averaging over both quantum and classical sources of randomness. For a given realization of waiting times, one first performs a quantum average over all possible measurement outcomes:
\begin{equation}
\langle \hat{O} \rangle_{\{\tau_i\}} =
\sum_{\boldsymbol{x}} \frac{\mathrm{Tr}\left[\hat{O}\, \tilde{\sigma}^{(\boldsymbol{x})}(t)\right]}{\mathrm{Tr}\left[\tilde{\sigma}^{(\boldsymbol{x})}(t)\right]}.
\end{equation}
Subsequently, one performs a classical average over the stochastic waiting times:
\begin{equation}
\langle \hat{O} \rangle =
\mathbb{E}_{\{\tau_i\}} \left[ \langle \hat{O} \rangle_{\{\tau_i\}} \right].
\end{equation}
Thus, the physically relevant expectation values are obtained via a combined averaging over quantum measurement outcomes (quantum trajectories) and over realizations of the Poisson-distributed measurement times.
\begin{figure}[t]
\centering
\includegraphics[width=0.7\linewidth]{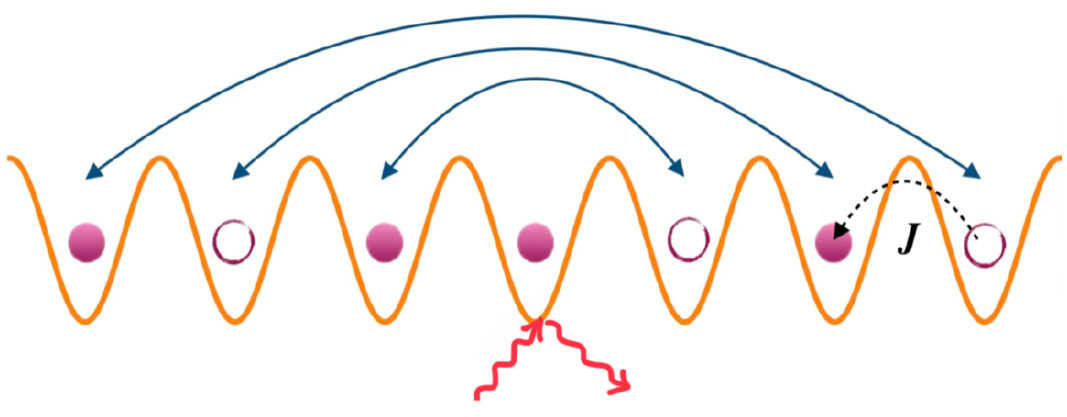}    
\caption{Schematic representation of the proposed setup for an open one-dimensional fermionic tight-binding lattice with nearest-neighbor hopping, subjected to stochastic projective measurements performed at the central site leading to an effective local dephasing dynamics described by Eq.~\eqref{eq:Dyn}.}
\label{fig:enter-label}
\end{figure}

\subsection*{3. Open Quantum Description of Stochastic Non-Selective Measurement}

We consider a one-dimensional fermionic tight-binding chain with an odd number of lattice sites $N$ and open boundary conditions. The system is governed by the Hamiltonian
\begin{equation}
\hat{H} = -J \sum_{i=1}^{N-1} \left( \hat{f}_{i}^{\dagger}\hat{f}_{i+1} + \hat{f}_{i+1}^{\dagger}\hat{f}_{i} \right),
\end{equation}
where $J$ denotes the nearest-neighbor tunneling amplitude, and $\hat{f}_{i}$ ($\hat{f}_{i}^{\dagger}$) is the fermionic annihilation (creation) operator at site $i$. Throughout this work, we set $J=\hbar=1$ for convenience. Such tight-binding models are routinely realized in a variety of quantum simulation platforms \cite{Georgescu14,Altman21}. We focus on the central lattice site $c = (N+1)/2$, where a detector is placed. The detector performs measurements at random times, with the waiting times between successive clicks distributed according to a Poisson process with rate $\gamma$.

Let $\hat{\rho}(t)$ denote the average density matrix of the system immediately before a detection event at time $t$. The measurement corresponds to probing the occupation of the central site via the projection operator
\begin{equation}
\hat{P} = \hat{f}_c^\dagger \hat{f}_c = \hat n_c.
\end{equation}
Upon measurement, the state undergoes a stochastic update according to
\begin{equation}
\lim_{\epsilon \to 0} \hat{\rho}(t+\epsilon) =
\begin{cases}
\dfrac{\hat{P}\hat{\rho}(t)\hat{P}}{\mathrm{Tr}[\hat{P}\hat{\rho}(t)]}, 
& \text{with probability } \mathrm{Tr}[\hat{P}\hat{\rho}(t)], \\[10pt]
\dfrac{\hat{P'}\hat{\rho}(t)\hat{P'}}{\mathrm{Tr}[\hat{P'}\hat{\rho}(t)]}, 
& \text{with probability } \mathrm{Tr}[\hat{P'}\hat{\rho}(t)].
\end{cases}
\end{equation}
where $\hat{P'}=\mathbb{I}-\hat{P}$ is the projector onto the rest of the lattice sites.

To derive the coarse-grained dynamics, we consider a short time interval $\Delta t$ such that $\gamma \Delta t \ll 1$. Within this interval, either no detection occurs with probability ($1 - \gamma \Delta t$) or a single detection occurs with probability $\gamma \Delta t$. In the absence of measurement, the system evolves unitarily under $\hat{H}$, yielding
\begin{equation}
\hat{\rho}(t+\Delta t) = \hat{\rho}(t) - i[\hat{H},\hat{\rho}(t)]\Delta t + \mathcal{O}(\Delta t^2).
\end{equation}
Including both unitary evolution and stochastic measurements, the density matrix evolves as
\begin{equation}
    \hat{\rho}(t+\Delta t) = (1 - \gamma \Delta t)\left[\hat{\rho}(t) - i[\hat{H},\hat{\rho}(t)]\Delta t \right] \nonumber + \gamma \Delta t \left( \hat{P}\hat{\rho}(t)\hat{P} + \hat{P'}\hat{\rho}(t)\hat{P'} \right)
+ \mathcal{O}(\Delta t^2).
\end{equation}

Retaining terms up to linear order in $\Delta t$ and taking the limit $\Delta t \to 0$, we obtain the differential equation
\begin{equation}
\frac{d\hat{\rho}}{dt}
= -i[\hat{H},\hat{\rho}]
+ \gamma \left( 2\hat{P}\hat{\rho}\hat{P} - \{\hat{P},\hat{\rho}\} \right).
\end{equation}
Defining the jump operator $\hat{L}_c \equiv \hat{P}$ and using the fermionic projector property $\hat{P}^\dagger \hat P = \hat n_c \hat n_c = \hat n_c = \hat{P}$, the above equation can be cast into the standard Lindblad form \cite{Lindblad76, Gorini76, Rivas10}:
\begin{equation}
\frac{d\hat{\rho}}{dt}
= -i[\hat{H},\hat{\rho}]
+ 2\gamma \left(
\hat{L}_c \hat{\rho} \hat{L}_c^\dagger
- \frac{1}{2}\{\hat{L}_c^\dagger \hat{L}_c,\hat{\rho}\}
\right)=\mathcal{L}(\hat\rho),
\label{eq:Dyn}
\end{equation}
which describes local dephasing induced by stochastic, non-selective measurements at the central site. 

The local dephasing dynamics considered here can be realized in two different ways that include direct stochastic measurements of the central-site occupation, which is discussed above, and noise-assisted implementation through a fluctuating light shift. There are available settings in which the local occupation can be monitored directly via an off-resonant beam addressed to the central site.  The scattered photons encode information regarding the occupancy of a site. In the strong-measurement limit, this approaches the projective protocol of Eqs.~(\ref{eq:one}-\ref{eq:three}). The connection between spontaneous photon scattering and decoherence in optical lattices is discussed, for example, in \cite{Zoller2010}. Nondestructive and spatially resolved measurement schemes for ultracold atoms are also discussed in \cite{Mekhov2012,Ueda2015}.

A noisy laser provides an alternate route that does not require performing measurement. A non-selective measurement and an appropriately engineered fluctuating light shift can generate the same dynamics. A measurement of $\hat n_c$, whose outcome is not recorded, implements the channel $\hat P\hat\rho\hat P+\hat P'\hat\rho\hat P'$. By contrast, a noisy laser produces continuous random-phase accumulation rather than discrete measurement jumps. For a far-detuned field, the AC Stark shift is proportional to the laser intensity \cite{Autler55}. Writing the laser intensity as $I(t)=I_0+\delta I(t)$,  its fluctuating part produces $\xi(t)\hat n_c$, giving rise to a time-dependent stochastic Hamiltonian, $\hat H_\xi(t)=\hat H+\xi(t)\hat n_c$. Treating the fluctuations as zero-mean Gaussian white noise yields
$\overline{\xi(t)}=0$ and
$\overline{\xi(t)\xi(t')}=2\Gamma\delta(t-t')$, where $\Gamma$ sets the strength of the light-shift fluctuations. Noise averaging suppresses the coherence between two sectors, the occupied central site and the empty central site sectors. Identifying  $\Gamma=\gamma$, the averaged density matrix generates the exact dynamical evolution governed by Eq.~\eqref{eq:Dyn}  \cite{Kubo63,Chenu17}.

For the noise-assisted route, key ingredients are available. Single-site optical addressing has been achieved using tightly focused beams \cite{Weitenberg11} and holographically shaped light \cite{Zupancic16}, while site-resolved detection is available through quantum-gas microscopy \cite{Gross21,Buob24}. The effects of intensity and fluctuations on far-detuned optical potentials are well established \cite{Savard97}, and light-shift-related dephasing has been measured \cite{Meschede05}. Controlled dephasing through engineered noise spectra has also been demonstrated in photonic systems \cite{Piilo18} and in superconducting qubits, where field fluctuations generate random AC Stark shifts \cite{Schoelkopf05}.

\subsection*{4. Strong Symmetry}

This central site dephasing dynamics turns out to obey the constraint of motion due to the conserved charge of the following symmetry operator:
\begin{equation}
    \hat{\mathcal{Q}}=\sum_{i=1}^{N} \hat{f}_{i}^{\dagger} \hat{f}_{N+1-i}.
\end{equation}
$\hat{\mathcal{Q}}$ is a strong symmetry operator, as it can be shown that $[\hat{\mathcal{Q}},\hat{H}]=0$ and $[\hat{\mathcal{Q}},\hat{L}_c]=0$, in contrast to the qubit array with pump and loss at the central site, for which $(\hat{\mathcal{Q}}-\frac{1}{2})^2$, instead of $\hat{\mathcal{Q}}$, is the strong symmetry of the dynamics \cite{Dutta20}.

To evaluate the commutator $[\hat{\mathcal{Q}},\hat H]$, we use the identity
\begin{equation}
[\hat f_a^\dagger \hat f_b,\hat f_c^\dagger \hat f_d]
= \delta_{bc}\,\hat f_a^\dagger \hat f_d
- \delta_{ad}\,\hat f_c^\dagger \hat f_b .
\label{eq:identity}
\end{equation}

\noindent We write the commutator explicitly,
\begin{equation}
[\hat{\mathcal{Q}},\hat H]
= -J \sum_{j=1}^{N}\sum_{i=1}^{N-1}
\Big([\hat f_j^\dagger \hat f_{N+1-j},\hat f_i^\dagger \hat f_{i+1}]
+[\hat f_j^\dagger \hat f_{N+1-j},\hat f_{i+1}^\dagger \hat f_i]
\Big).
\end{equation}

\noindent Evaluating the two commutators yields
\begin{align}
[\hat f_j^\dagger \hat f_{N+1-j},\hat f_i^\dagger \hat f_{i+1}]
&= \delta_{N+1-j,i}\,\hat f_j^\dagger \hat f_{i+1}
- \delta_{j,i+1}\,\hat f_i^\dagger \hat f_{N+1-j}, \\
[\hat f_j^\dagger \hat f_{N+1-j},\hat f_{i+1}^\dagger \hat f_i]
&= \delta_{N+1-j,i+1}\,\hat f_j^\dagger \hat f_i
- \delta_{j,i}\,\hat f_{i+1}^\dagger \hat f_{N+1-j}.
\end{align}

\noindent Performing the sum over \(j\), we obtain
\begin{equation}
[\hat{\mathcal{Q}},\hat H]
= -J \sum_{i=1}^{N-1}
\Big(\hat f_{N+1-i}^\dagger \hat f_{i+1}
- \hat f_i^\dagger \hat f_{N-i} + \hat f_{N-i}^\dagger \hat f_i
- \hat f_{i+1}^\dagger \hat f_{N+1-i}
\Big).
\end{equation}

\noindent The terms cancel pairwise upon relabelling indices, and we finally arrive at $[\hat{\mathcal{Q}},\hat H]=0$.

To evaluate the commutator $[\hat{\mathcal{Q}},\hat L_c]=[\hat{\mathcal{Q}},\hat n_c]$, we again use the identity \eqref{eq:identity},
\begin{align}
[\hat {\mathcal{Q}},\hat n_c]
&= \sum_{j=1}^{L}
[\hat f_j^\dagger \hat f_{N+1-j},\hat f_c^\dagger \hat f_c] \\
&= \sum_{j=1}^{N}
\left(
\delta_{N+1-j,c}\,\hat f_j^\dagger \hat f_c
- \delta_{j,c}\,\hat f_c^\dagger \hat f_{N+1-j}
\right).
\end{align}
For odd \(N\), one has \(N+1-c=c\), and therefore both terms contribute only for \(j=c\), yielding
\begin{equation}
[\hat {\mathcal{Q}},\hat n_c]
= \hat f_c^\dagger \hat f_c - \hat f_c^\dagger \hat f_c = 0 .
\end{equation}

This strong symmetry operator can also be re-expressed as
\begin{equation}
\hat{\mathcal{Q}}=\hat{\nu}_{\mathrm{e}}-\hat{\nu}_{\mathrm{o}},
\label{eq:Q_new}
\end{equation}
where $\hat{\nu}_{\mathrm{e}}$ and $\hat{\nu}_{\mathrm{o}}$ are, respectively, the total occupation of the single-particle even-parity and odd-parity states (see \ref{Appendix_A} for the derivation).

For $\mathcal{N}_{f}$ non-interacting fermions on a lattice of odd size $N$, the expression in Eq.~\eqref{eq:Q_new} provides a direct interpretation of the eigenvalues of $\hat{\mathcal{Q}}$. In particular, if a many-body state $|\Psi\rangle_{\mathcal{N}_{\!f}}$ contains $\nu_{\mathrm{e}}$ fermions in the even-parity single-particle states and $\nu_{\mathrm{o}}$ fermions in the odd-parity single-particle states, then $\hat{\mathcal{Q}}|\Psi\rangle_{\mathcal{N}_{\!f}}=\lambda |\Psi\rangle_{\mathcal{N}_{\!f}}$ where $\lambda=\nu_{\mathrm{e}}-\nu_{\mathrm{o}}$. We refer to $\lambda$ as the \emph{charge} of the corresponding symmetry sector.

For an odd lattice, there are $(N+1)/2$ even-parity and $(N-1)/2$
odd-parity single-particle states. Consequently, the occupations satisfy $0\leq\nu_{\mathrm{e}}\leq\frac{N+1}{2}$ and $0\leq\nu_{\mathrm{o}}\leq\frac{N-1}{2}$. Furthermore, since the dynamics conserves the total particle number,
the occupations are constrained by $\nu_{\mathrm{e}}+\nu_{\mathrm{o}}=\mathcal{N}_{f}$. It follows that the charge can equivalently be written as $    \lambda=2\nu_{\mathrm{e}}-\mathcal{N}_{f}=\mathcal{N}_{f}-2\nu_{\mathrm{o}}$. Thus, for a fixed particle number, specifying the charge $\lambda$
uniquely determines the number of particles in the even- and odd-parity
sectors.

The strong symmetry therefore induces a decomposition of the many-body
Hilbert space into invariant charge sectors, $\mathcal{H}=\bigoplus_{\lambda}\mathcal{H}_{\lambda},$ where $\mathcal{H}_{\lambda}$ denotes the subspace containing all states with a fixed value of the charge $\lambda$. Since $[\hat{\mathcal{Q}},\hat{H}]=0,\;[\hat{\mathcal{Q}},\hat{L}_{c}]=0,$ neither the coherent evolution nor the central-site dephasing can couple states belonging to different charge sectors. The corresponding Lindbladian consequently preserves this decomposition. For a given pair $(\nu_{\mathrm{e}},\nu_{\mathrm{o}})$ satisfying
$\nu_{\mathrm{e}}+\nu_{\mathrm{o}}=\mathcal{N}_{f}$, the dimension of the
corresponding charge sector is $\dim\mathcal{H}_{\lambda}
=\dbinom{(N+1)/2}{\nu_{\mathrm{e}}}
\dbinom{(N-1)/2}{\nu_{\mathrm{o}}}.$ The first factor counts the ways of choosing $\nu_{\mathrm{e}}$ occupied
states from the $(N+1)/2$ even-parity single-particle states, while the
second factor counts the ways of choosing $\nu_{\mathrm{o}}$ occupied
states from the $(N-1)/2$ odd-parity states.

\subsection*{5. Unique Single Particle Steady State}
When we consider a single particle, the operator $\hat{\mathcal{Q}}$ has only two eigenvalues, $+1$ and $-1$. The $+1$ sector has $(N+1)/2$ eigenstates and the $-1$ sector has $(N-1)/2$ eigenstates. These are nothing but the single-particle even and odd parity sectors. Therefore, in the case of single-particle we can write $ \mathcal{H} = \mathcal{H}_+  \oplus \mathcal{H}_-$, where $\mathcal{H}_+$ spans the even-parity sector of the total Hilbert space $\mathcal{H}$ having $\dim(\mathcal{H}_+)=\dfrac{N+1}{2}$ and $\mathcal{H}_-$ spans the odd-parity sector having $\dim(\mathcal{H}_-)=\dfrac{N-1}{2}$. As $\hat{\mathcal{Q}}$ also commutes with $\hat{L}_c$, the dynamics is constrained within the same symmetry sector of $\hat{\mathcal{Q}}$ for all time. However, the odd parity states are not affected by the dephasing mechanism, which we show below.

Let $\{|\psi_\alpha\rangle\}$ be the complete set of single-particle eigenstates of $\hat H$ with energy $\{\varepsilon_\alpha\}$:
\begin{equation}
\hat H|\psi_\alpha\rangle=\varepsilon_\alpha|\psi_\alpha\rangle.
\end{equation}
Now for the open boundary conditions, the eigenfunctions are given by
\begin{equation}
\psi_{\alpha}(j)
= \langle j |\psi_\alpha \rangle =
\sqrt{\frac{2}{N+1}}
\sin\!\left(\frac{\pi (\alpha+1) j}{N+1}\right).
\end{equation}
where $j$ and $\alpha$ correspond to the site index and eigenstate index, respectively. $\alpha$ takes values $0,1,2,\dots,N-1$. One can easily check that even $\alpha$ values correspond to even-parity states and odd $\alpha$ values correspond to odd-parity states. For the central-site $c=(N+1)/2$, for a specific energy state $|\psi_\alpha\rangle$, the overlap $\phi_\alpha$ turns out to be, 
\begin{equation}
\phi_{\alpha}\equiv\psi_{\alpha}(c)=\langle c| \psi_\alpha \rangle = \sqrt{\frac{2}{N+1}}
\sin\!\left(\frac{\pi (\alpha+1)}{2}\right)
\end{equation}
which is $\pm \sqrt{\dfrac{2}{N+1}}$ for even $\alpha$ and zero for odd $\alpha$. We therefore define bright and dark subspaces \cite{Buca12} corresponding to even-parity (even $\alpha$ modes) and odd-parity (odd $\alpha$ modes) as follows: $ \mathcal{H}_+=\{|\psi_\alpha\rangle : \phi^2_\alpha=\frac{2}{N+1}\}\; \text{and} \; \mathcal{H}_-=\{|\psi_\alpha\rangle : \phi^2_\alpha=0\}$ respectively. States in $\mathcal{H}_-$ possess a node at the central site and hence are completely unaffected by the Lindblad operator. Any state inside $\mathcal{H}_-$ therefore evolves purely unitarily. We therefore restrict our attention to the bright subspace $\mathcal{H}_+$ on which the dephasing mechanism has a non-trivial effect.

Next, we investigate whether there is any unique steady state in the bright sector. For the uniqueness, we will use the Frigerio-Spohn theorem \cite{Spohn77,Frigerio78} which says: If the only operator commuting with both the Hamiltonian $\hat{H}$ and all the Lindblad operators $\hat{L}_i$ is a multiple of the identity, then the dynamics is ergodic, and there exists a unique steady state $\hat\rho^{\text{ss}}$ (superscript ss is for steady-state), to which the density matrix converges at long times for any initial condition,
\begin{equation}
\lim_{t\to\infty}e^{\hat{\mathcal{L}}t}\hat\rho(0)=\hat\rho^{\text{ss}}.
\end{equation}

To invoke this theorem and establish the existence of a unique steady state in the even-parity sector, it is necessary and sufficient to show that any operator \(\hat A\) satisfying
\begin{equation}
[\hat A,\hat H_+] = [\hat A,\hat L_c]=0
\end{equation}
must be proportional to the identity operator on the even-parity subspace. The detailed proof of the uniqueness of the steady state in the even-parity sector is given in \ref{Appendix_B}. In the even-parity sector the corresponding steady-state density matrix \(\hat\rho^{\mathrm{ss}}\) satisfies the Lindblad equation
\begin{equation}
\mathcal{L}(\hat\rho^{\mathrm{ss}})=-i[\hat H_+,\hat\rho^{\mathrm{ss}}]
+
2\gamma
\left(
\hat L_c \hat\rho^{\mathrm{ss}} \hat L_c^\dagger
-
\frac12
\left\{
\hat L_c^\dagger \hat L_c,
\hat\rho^{\mathrm{ss}}
\right\}
\right)
=0.
\label{eq:Lindblad_H_plus}
\end{equation}

\begin{figure}[t]
    \centering
    \includegraphics[width=0.9\linewidth]{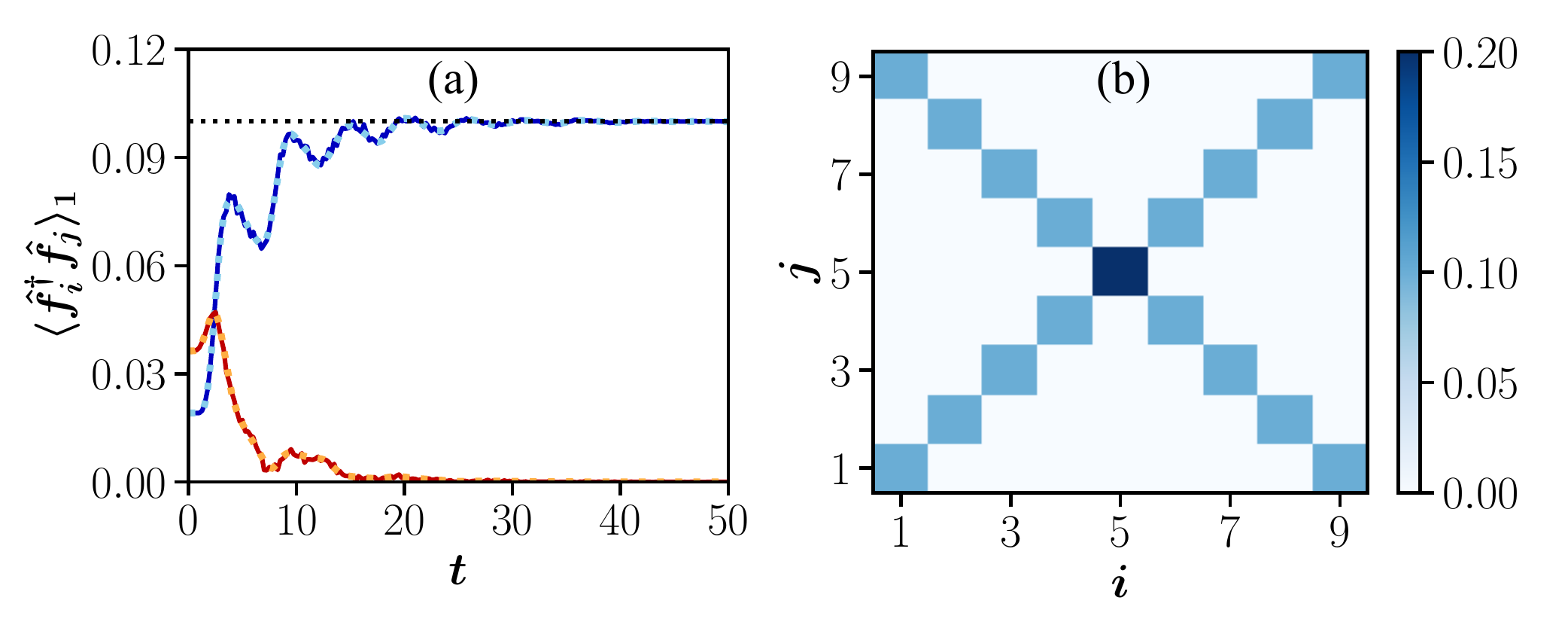}
    \caption{ (a) depicts the dynamical evolution of the end-to-end correlation, $\langle \hat{f}_1^\dagger \hat{f}_N \rangle_1$ (the blue solid line corresponds to the stochastic result and the sky-blue dotted line corresponds to the result obtained by Lindbladian dynamics), and the correlation between the first and the second lattice sites, $\langle \hat{f}_1^\dagger \hat{f}_2 \rangle_1$ (the red solid line corresponds to the stochastic result and the orange dotted line corresponds to the result obtained by Lindbladian dynamics) for a system size of $N=9$. For the stochastic result, an ensemble size of 5000 is used. The black dotted line corresponds to the steady-state value of $1/10$. (b) shows the corresponding steady-state correlation matrix $\langle \hat{f}_i^\dagger \hat{f}_j \rangle^{\text{ss}}_1$. For both the sub-figures, the system is initiated in the single-particle ground state of $\hat{H}$- an even-parity state, and we have taken $\gamma=0.5$.}
    \label{fig:Dephasing}
\end{figure}

Since the jump operator is a projection operator, it is self-adjoint, \(L_c^\dagger = L_c\). Consequently, the corresponding Lindbladian is unital, implying that the identity operator is invariant under the Lindblad evolution, i.e., \(\mathcal{L}(\hat{\mathbb I}_+)=0\). Therefore,
\begin{equation}
    \hat{\rho}^\text{ss}=k\,\mathbb{I_+},
    \label{eq:ss_sol}
\end{equation}
where $k=1/\Tr{\mathbb{I_+}}$ which preserves the trace of $\hat{\rho}^\text{ss}$ to be unity. As the even sector has a dimension $\dfrac{N+1}{2}$, we get $k=\dfrac{2}{N+1}$. Hence,
\begin{equation}
    \hat{\rho}^\text{ss}=\frac{2}{N+1}\mathbb{I_+}.
    \label{eq:ss_sol}
\end{equation}
Writing $\mathbb{I_+}$ in the lattice basis, the unique single-particle steady-state density matrix turns out to be
\begin{equation}
    \hat{\rho}^{\text{ss}} = \frac{1}{(N+1)} \sum_{i} \left( |i\rangle \langle i| + |i\rangle \langle N+1-i| \right).
\end{equation}

Fig.~\ref{fig:Dephasing}(a) shows the time evolution of the correlation between two adjacent lattice sites and two distant lattice sites, initiated from the ground state of a single particle in a lattice of size $N=9$. From this figure, it is evident that long-range correlated pairs are generated in the long-time evolved unique steady state. Fig.~\ref{fig:Dephasing}(b) depicts the corresponding dynamically obtained steady-state correlation matrix $\langle \hat{f}_i^{\dagger}\hat{f}_j \rangle_{1}^{\text{ss}}$, which is nothing but the steady-state density matrix $\hat{\rho}^{\text{ss}}$ for a single particle. $\langle \cdot \rangle_1$ denotes the expectation value of an operator in a single-particle state. In \ref{Appendix_C} we have given an explicit analytical calculation of the dephasing dynamics for the simplest case $N=3$.

\subsection*{6. Pairwise Mode Entanglement in the Single-Particle Steady State} 

In this section, we provide a way to determine whether the symmetrically located correlated lattice sites are entangled or not, and if they are entangled, how to quantify that. For the determination of whether they are entangled or not, we will use the Peres–Horodecki criterion or the PPT criterion \cite{Peres96, Horodecki77}, which provides a necessary condition for the joint density matrix $\rho_{AB}$ of two parties, $A$ and $B$, to be separable, which also turns out to be a sufficient condition in the $2 \times 2$ and $2 \times 3$ dimensional cases. 

\begin{figure}[t]
    \centering
    \includegraphics[width=0.9\linewidth]{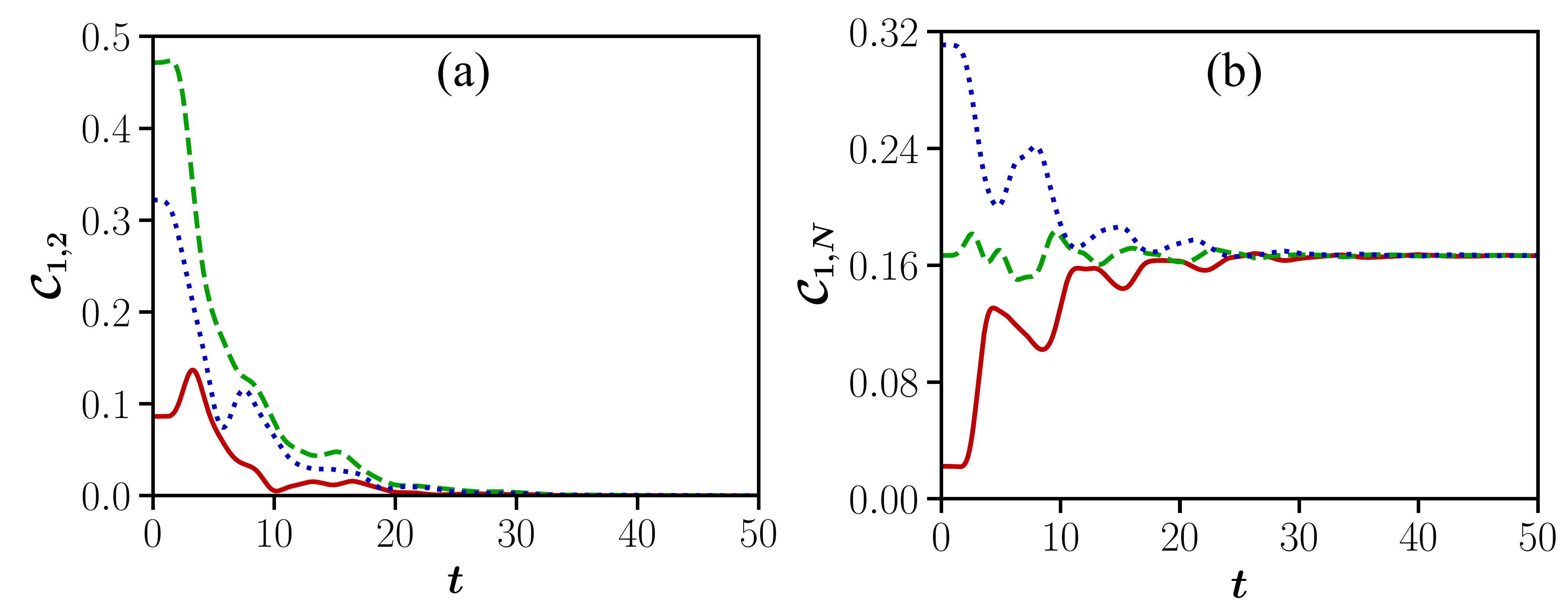}
    \caption{shows the time variation of first-to-second site concurrence $\mathcal{C}_{1,2}$ (panel (a)) and end-to-end site concurrence $\mathcal{C}_{1,N}$ (panel (b)) for a single particle, for system size $N=11$, and demonstrates generation of pairwise mode entanglement for different initial states: ground state (red solid line), 2nd excited state (green dashed line), and 4th excited state (blue dotted line). We have taken $\gamma=0.5$ for all the cases.}
    \label{fig:Conc vs Time}
\end{figure}

The steady-state density matrix has matrix elements,
\begin{equation}
\rho^{\text{ss}}_{ij}=\langle \hat{f}_i^{\dagger}\hat{f}_j \rangle_1^{\text{ss}}=\frac{1}{N+1}[\delta_{ij}+\delta_{i(N+1-j)}].
\end{equation}
Therefore, in the steady state, the generic form of the single-particle reduced density matrix between sites $i$ and $j=(N+1-i)$ is found to be,
\begin{equation}
    \hat\rho^{\text{ss}}_{\text{rd}}=
    \begin{bmatrix}
    \dfrac{N-1}{N+1} & 0 & 0 & 0\\
    0 & \dfrac{1}{N+1} & \dfrac{1}{N+1} & 0\\[8pt] 
    0 & \dfrac{1}{N+1} & \dfrac{1}{N+1} & 0\\
    0 & 0 & 0 & 0
    \end{bmatrix},
    \label{eq:red_ss_den_mat}
\end{equation}
which, under the partial transposition map, turns out to be
\begin{align}
 [\hat{\rho}^{\text{ss}}_{\text{rd}}]^{T} = \left(\mathbb{I} \otimes T\right) \hat\rho^{\text{ss}}_{\text{rd}} =
 \begin{bmatrix}
    \dfrac{N-1}{N+1} & 0 & 0 & \dfrac{1}{N+1} \\
    0 & \dfrac{1}{N+1} & 0 & 0\\
    0 & 0 & \dfrac{1}{N+1} & 0\\
    \dfrac{1}{N+1} & 0 & 0 & 0
\end{bmatrix}.
\end{align}
The eigenvalues of $[\hat{\rho}^{\text{ss}}_{\text{rd}}]^{T}$ are given by $\frac{1}{N+1},\frac{1}{N+1},\frac{-1+N+\sqrt{5-2 N+N^2}}{2(1+N)},$ and $\frac{-1+N-\sqrt{5-2 N+N^2}}{2(1+N)}$. As it can be seen that the last eigenvalue is negative for any arbitrary $N$, and has a leading-order behavior of $-(1/N^2)$ in the large $N$ limit, the symmetrically correlated pairs represented by the reduced state, $\hat{\rho}^{\text{ss}}_{\text{rd}}$, are guaranteed to be entangled in a finite-size lattice. Here, the sites $i$ and $j=(N+1-i)$ are entangled, meaning coherent delocalization between the two product states $|10\rangle_{ij}=|1\rangle_i\otimes|0\rangle_j$ and $|01\rangle_{ij}=|0\rangle_i\otimes|1\rangle_j$. This pairwise fermionic-mode entanglement between the occupation modes of two mirror-related sites is quantified by concurrence. For two modes \(i\) and \(j\), the concurrence \(\mathcal C_{ij}\) quantifies their pairwise quantum entanglement, with \(\mathcal C_{ij}=0\) for separable modes and \(\mathcal C_{ij}>0\) indicating nonzero entanglement. For fermionic modes, this entanglement refers to the quantum correlations between their occupation degrees of freedom and is distinct from particle-particle entanglement. The concurrence satisfies \(0\leq \mathcal C_{ij}\leq1\), with \(\mathcal C_{ij}=1\) corresponding to maximal pairwise mode entanglement. The concurrence between two sites of a many-body lattice has a long history as a diagnostic of quantum phenomena. {\color{black} It was introduced in this context to characterize quantum critical points in spin chains \cite{Osterloh02}, and was subsequently extended to a wide range of spin, Hubbard, and Hubbard-type models \cite{Sadhukhan, Mishra_spin_glass, Mishra16, Mishra19, Deng05, Sousa18, Abaach21, Abaach23, Roosz24, Bippus26} and the framework reviewed in \cite{Enk, Heaney, Killoran}.} The pairwise mode entanglement is a useful resource in quantum communication protocols, teleportation, quantum cryptography, {\color{black} and in the investigation of violating Bell inequalities} \cite{Heaney, Killoran,Morris,Fuwa}.

We use the standard definition of concurrence, $\mathcal{C}_{ij}$, between lattice sites $i$ and $j$ \cite{Hill97, Wootters98, Horodecki09}. The definition of concurrence is $\mathcal{C}_{ij}(\hat\rho_{\text{rd}})\equiv \text{max}(0,\Lambda_{1}-\Lambda_{2}-\Lambda_{3}-\Lambda_{4})$ where $\hat\rho_{\text{rd}}$ is the steady state reduced density matrix between lattice sites $i$ and $j$; $\Lambda_{1},\Lambda_{2},\Lambda_{3},\Lambda_{4}$ are the eigenvalues of the matrix $\mathcal{M}=\sqrt{\sqrt{\hat\rho_{\text{rd}}}\, \hat{\tilde{\rho}}_{\text{rd}} \sqrt{\hat\rho_{\text{rd}}}}$ in descending order and $\hat{\tilde{\rho}}_{\text{rd}}=(\sigma_y\otimes\sigma_y)\hat\rho_{\text{rd}}(\sigma_y\otimes\sigma_y),\, \sigma_y$ is the usual Pauli matrix. Using $\hat\rho_{\text{rd}}^{\rm ss}$ as in Eq.~\eqref{eq:red_ss_den_mat}, the eigenvalues of the $\mathcal{M}$ matrix turn out to be ${2}/{(N+1)},0,0,0$, which gives the steady-state concurrence for a single particle to be:
\begin{equation}
    \mathcal{C}^{\rm ss}_{i,N+1-i} = \frac{2}{(N+1)}.
\end{equation}

In Fig.~\ref{fig:Conc vs Time}, we present the time variation of the end-to-end concurrence $\mathcal{C}_{1,N}$ for a single particle for a system size of $N=11$, starting from three different initial states. We find that each of them converges towards a steady value of $1/6$, as expected. For excited states, the initial end-to-end mode entanglement is higher than that of the steady-state value. However, initially, there is also non-zero mode entanglement between all other sites, which disappears in the steady state, and the mode entanglement remains only between pairs located symmetrically.

\begin{figure}[t]
    \centering
    \includegraphics[width=0.5\linewidth]{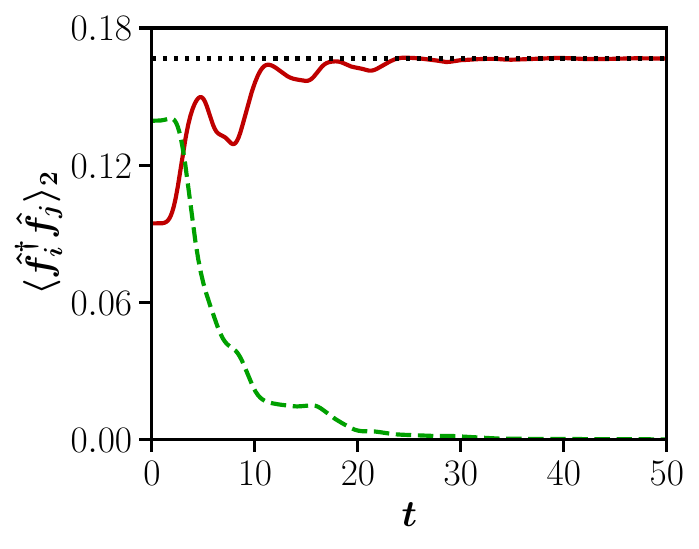}
    \caption{illustrates the temporal evolution of the end-to-end correlation, $\langle \hat{f}_1^{\dagger}\hat{f}_{N} \rangle_2$ (red solid line), and the nearest-neighbor correlation, $\langle \hat{f}_1^{\dagger}\hat{f}_{2} \rangle_2$ (green dashed line), for a two-particle system ($\mathcal{N}_{\!f}=2$) with system size $N=11$, initialized in its first excited state. The end-to-end correlation $\langle \hat{f}_1^{\dagger}\hat{f}_{N} \rangle_2$ asymptotically approaches the analytically predicted saturation value $\mathcal{N}_{\!f}/(N+1)=1/6$ (black dotted line). In the non-interacting limit, the fermionic two-particle first excited state is constructed by proper anti-symmetrization of the single-particle ground state and second excited state, leading to the emergence of a long-range entangled steady state. Here also $\gamma=0.5$.}
    \label{fig:L11_N2}
\end{figure}

\subsection*{7. Multi-Particle Case}  
Here, the model we consider is a non-interacting fermionic model with local dephasing. So, any multi-particle state can be constructed using the single-particle states with proper anti-symmetrization due to the Pauli exclusion principle. For a single particle, we have seen that the dephasing mechanism affects only the even-parity states and thus only the initial states having even-parity go to the steady-state that gives rise to long-range pairwise mode entanglement. So, for the multi-particle case, if the initial state is composed of only even-parity single-particle states, we get a steady state. 

\begin{figure}[t]
    \centering
    \includegraphics[width=0.7\linewidth]{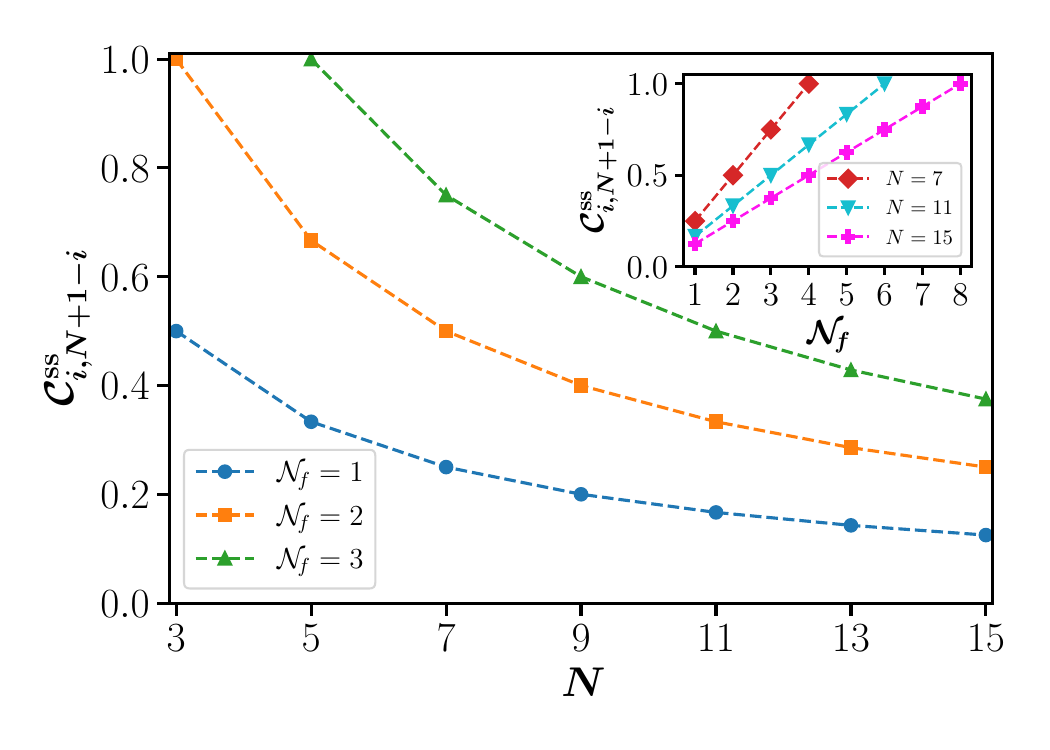}
    \caption{The main panel depicts the steady-state concurrence $\mathcal{C}^{\rm ss}_{i,N+1-i}$ between symmetrically located sites $i$ and $N+1-i$ as a function of the system size $N$ for different particle fillings $\mathcal{N}_{\!f}$. The concurrence exhibits an inverse scaling with the system size, $\mathcal{C}^{\rm ss}_{i,N+1-i}\sim 1/N$, whereas, for a fixed system size $N$, it increases linearly with the particle filling $\mathcal{N}_{\!f}$, as shown in the inset. The numerical data are in excellent agreement with the analytical expression given in Eq.~\eqref{eq:Conc_ss}, confirming the validity of the derived steady-state concurrence. The dashed lines are guides to the eye.}
    \label{fig:Conc_ss}
\end{figure}

For a lattice of size $N$, where $N$ is odd, there are $(N+1)/2$ even-parity single-particle eigenstates. Hence, if there are $\mathcal{N}_{\!f}$ fermions, there will be $\dbinom{(N+1)/2}{\mathcal{N}_{\!f}}$ linearly independent many-particle eigenstates, where $\mathcal{N}_{\!f}\le (N+1)/2$, for which the steady state turns out to be the one having long-ranged entangled pairs. If the dynamics is initialized in an antisymmetric ${\mathcal{N}_{\!f}}$-fermion state whose occupied single-particle eigenstates all belong to the bright subspace, the system relaxes to the maximally mixed state within the corresponding ${\mathcal{N}_{\!f}}$-particle bright sector. Equivalently, the steady-state density matrix is proportional to the projector onto the antisymmetric ${\mathcal{N}_{\!f}}$-fermion subspace of the bright sector. Thus, if we consider the one-body fermionic correlator, the steady-state correlators between the sites $i$ and $j$ in the case of $\mathcal{N}_{\!f}$ non-interacting fermions are given by
\begin{equation}
    \langle \hat{f}_i^{\dagger}\hat{f}_{j} \rangle _{\mathcal{N}_{\!f}} ^{\text{ss}} =\mathcal{N}_{\!f}\langle \hat{f}_i^{\dagger}\hat{f_j} \rangle_1^{\text{ss}}= \frac{\mathcal{N}_{\!f}}{N+1}[\delta_{ij}+\delta_{i(N+1-j)}].
    \label{eq:N-ssd}
\end{equation}
From the above expression, the steady-state correlation between the symmetrically located sites $i$ and $N+1-i$ increases linearly with the particle filling $\mathcal{N}_{\!f}$ and decreases inversely with the system size $N$. This scaling is directly reflected in the steady-state concurrence between the corresponding sites, as discussed below.

In general, in a particle number conserving system, the reduced density matrix between site $i$ and site $j$ can be written in terms of $\hat n_i$, $\hat n_j$, and $\hat f_i^\dagger \hat f_j$ as follows,
\begin{equation}
\hat{\rho}_{\mathrm{rd}}(i,j)=
\begin{bmatrix}
1-\langle \hat n_i\rangle-\langle \hat n_j\rangle+\langle \hat n_i \hat n_j\rangle
&0&0&0\\
0&\langle \hat n_j\rangle-\langle \hat n_i \hat n_j\rangle
&\langle \hat f_j^\dagger \hat f_i\rangle&0\\
0&\langle \hat f_i^\dagger \hat f_j\rangle
&\langle n_i\rangle-\langle \hat n_i \hat n_j\rangle&0\\
0&0&0&\langle \hat n_i \hat n_j\rangle
\end{bmatrix}_{\!\mathcal{N}_{\!f}},
\label{eq:reduced}
\end{equation}
where the subscript ${\mathcal{N}_{\!f}}$ outside the matrix is used to emphasize that the expectation values are evaluated in the Hilbert space of ${\mathcal{N}_{\!f}}$ fermions.
See \ref{Appendix_D} for the detailed derivation of the reduced density matrix. The many-body fermionic state being Gaussian, we use the relation $\langle \hat n_i \hat n_j \rangle_{\mathcal{N}_{\!f}}=\langle \hat n_i \rangle_{\mathcal{N}_{\!f}} \langle \hat n_j \rangle_{\mathcal{N}_{\!f}} - |\langle \hat f_i^\dagger \hat f_j \rangle_{\mathcal{N}_{\!f}}|^2$. Eq.~\eqref{eq:N-ssd} gives $\langle \hat n_i \rangle_{\mathcal{N}_{\!f}} ^{\text{ss}}=\langle \hat n_j \rangle_{\mathcal{N}_{\!f}} ^{\text{ss}}=\dfrac{\mathcal{N}_{\!f}}{N+1}$ for $i,j\ne c$ and $\langle \hat f_i^\dagger \hat f_j \rangle ^{\text{ss}}_{\mathcal{N}_{\!f}}=\dfrac{\mathcal{N}_{\!f}}{N+1}$ for $j=(N+1-i)$. Therefore, the steady-state reduced density matrix between sites $i$ and $j=(N+1-i)$ becomes
\begin{align}
\hat\rho^{\text{ss}}_{\text{rd}}(i,N+1-i)
&=
\begin{bmatrix}
1 - \dfrac{2\mathcal{N}_{\!f}}{N+1} & 0 & 0 & 0 \\[4pt]
0 & \dfrac{\mathcal{N}_{\!f}}{N+1} & \dfrac{\mathcal{N}_{\!f}}{N+1} & 0 \\[8pt]
0 & \dfrac{\mathcal{N}_{\!f}}{N+1} & \dfrac{\mathcal{N}_{\!f}}{N+1} & 0 \\[4pt]
0 & 0 & 0 & 0
\end{bmatrix} \nonumber
\\ 
&=
\left(1 - \frac{2\mathcal{N}_{\!f}}{N+1}\right)
|00\rangle_{ij}\langle00|_{ij}
+\frac{2\mathcal{N}_{\!f}}{N+1}
|\chi_+\rangle_{ij} \langle \chi_+|_{ij} ,
\end{align}
where $|00\rangle_{ij}$ is the vacuum state and $|\chi_+\rangle_{ij}$ is the positively charged Bell state between sites $i$ and $j=(N+1-i)$ and $|\chi_+\rangle$ is given as
\begin{equation}
    |\chi_+\rangle = \frac{1}{\sqrt{2}}\left(|01\rangle + |10\rangle \right).
\end{equation}

This analytical steady-state form is verified by direct numerical evolution of the Lindblad equation for finite chains and different particle numbers. The concurrence for the above reduced density matrix comes out to be
\begin{equation}
    \mathcal{C}^{\rm ss}_{i,N+1-i} = \frac{2\mathcal{N}_{\!f}}{(N+1)}.
    \label{eq:Conc_ss}
\end{equation}
So for the multiparticle case, similar two-mode coherence between $|10\rangle_{ij}$ and $|01\rangle_{ij}$ survives with a particle-number-dependent weight $\mathcal{N}_{\!f}$, giving rise to enhanced entanglement between sites $i$ and $j=(N+1-i)$. The dependence of the steady-state concurrence on the system size and
particle filling reveals two distinct scaling behaviors. For a fixed
particle filling $\mathcal{N}_{\!f}$, the concurrence between the
mirror-symmetric sites $i$ and $N+1-i$ decreases with increasing
system size and scales as
$\mathcal{C}^{\rm ss}_{i,N+1-i}\sim 1/N$. In contrast, at a fixed system
size $N$, the steady-state concurrence increases
linearly with the particle filling $\mathcal{N}_{\!f}$, as shown in the inset of Fig.~\ref{fig:Conc_ss}. Writing $\omega=\mathcal{N}_{\!f}/N < 1/2$, $\lim_{N\to\infty} \mathcal{C}_{i,N+1-i}^{\rm ss} \to 2\omega$, so the pairwise mode entanglement survives the thermodynamic limit and is set by the filling, not by the separation. As $\omega\to 1/2$, $\mathcal{C}_{i,N+1-i}^{\rm ss} \to 1$. Entanglement becomes vanishing only in the dilute system at low density.

\begin{figure}[t]
   \centering
   \includegraphics[width=0.9\linewidth]{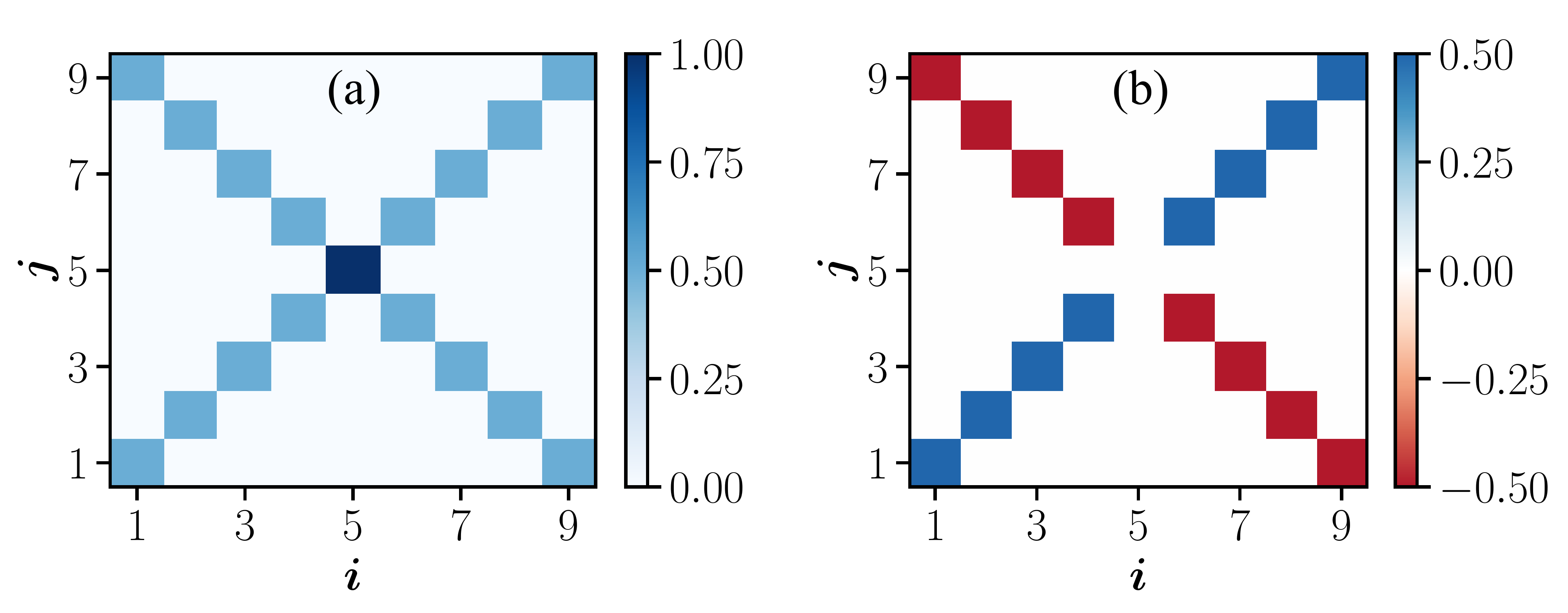}
   \caption{(a) and (b) show the correlation matrix $\langle \hat f_i^\dagger \hat f_j \rangle$ for the dark states of a system size of $N=9$, corresponding to $\nu_e=\mathcal{N}_{\!f}=5$ and $\nu_o=\mathcal{N}_{\!f}=4$, respectively, with the dynamics in each case initiated from the unique Hamiltonian eigenstate in the corresponding particle-number sector.} 
   \label{fig:Dark_State}
\end{figure}

Concurrence is a nonlinear functional of the reduced density matrix and is therefore not directly measurable as an observable. However, the two-site reduced density matrix in Eq.~\eqref{eq:reduced} is completely determined by the local densities $\langle \hat n_i \rangle_{\mathcal{N}_{\!f}}, \langle \hat n_j \rangle_{\mathcal{N}_{\!f}}$ and the coherence $\langle \hat f_i^\dagger \hat f_j\rangle_{\mathcal{N}_{\!f}}$. Site-resolved densities are directly accessible, while recent optical-lattice experiments have demonstrated access to off-diagonal correlations \cite{Impertro24}. The two-site reduced density matrix has to be reconstructed from these quantities, from which the concurrence can be determined.

An interesting scenario occurs when we have $\mathcal{N}_{\!f}=(N+1)/2$. It is clear from the eigensectors of $\hat {\mathcal{Q}}$ that there is only one eigenstate that corresponds to $\nu_\text{e} =\mathcal{N}_{\!f}=(N+1)/2$, the case for which all the fermions occupy all the available single-particle even-parity eigenstates of $\hat{H}$. As $\hat H$ and $\hat L_c$ commute with $\hat {\mathcal{Q}}$, this particular eigenstate turns out to be the steady-state solution of the Lindblad equation, Eq.~\eqref{eq:Dyn} and as a result it has the same `X' structure in the correlation matrix (see Fig.~\ref{fig:Dark_State}(a)) for all time. This state is also a so-called dark state; however, it is special in the sense that it remains unaffected even in the presence of dephasing. This is in contrast to the usual dark states, which are constructed from odd-parity eigenstates and are never exposed to the central-site dephasing mechanism. As this state supports long-range maximal pairwise mode entanglement, it is of definite interest in quantum information processing and quantum computing \cite{Lidar03,Kohout08}. Another interesting scenario is $\nu_\text{o}=\mathcal{N}_{\!f}=(N-1)/2$, i.e., all the single-particle odd-parity eigenstates are occupied. This is also a single state in the respective symmetry sector of $\hat {\mathcal{Q}}$ and hence it is indeed a steady state solution of Eq.~\eqref{eq:Dyn}. For this state, we also get long-range pairwise mode entanglement, but the correlation between the lattice pairs turns out to be the negative of that of the previous case (see Fig.~\ref{fig:Dark_State}(b)), and hence we get a maximally entangled negatively charged Bell pair between symmetrically located lattice sites:
\begin{equation}
    |\chi_-\rangle = \frac{1}{\sqrt{2}}\left(|01\rangle - |10\rangle \right).
\end{equation}
For these two cases $\nu_\text{e} =\mathcal{N}_{\!f}=(N+1)/2$ and $\nu_\text{o} =\mathcal{N}_{\!f}=(N-1)/2$, as the individual eigenstates do not evolve, the states always remain pure and, as a result, by tracing over the other sites we get pure states which are maximally entangled.

\subsection*{8. Conclusion}
In this work, we have explored the emergence of robust pairwise mode entanglement in an odd-sized open fermionic tight-binding chain subjected to stochastic projective measurements at its central site, effectively realizing a minimal local dephasing protocol. We have shown that, despite the simplicity of this mechanism, the interplay between measurement-induced dynamics and underlying symmetries gives rise to highly nontrivial steady-state behavior.

In the single-particle regime, we demonstrated that when the system is initialized in an even-parity state, the dynamics drive it toward a well-defined steady state, which we characterized analytically and corroborated numerically. This steady state exhibits a striking structure: it supports pairwise mode entanglement between symmetrically positioned sites across the lattice, even at arbitrarily large separations. This establishes local dephasing, typically associated with decoherence, as a resource for engineering long-range quantum correlations.

Extending our analysis to the many-particle case, we found that the degree of pairwise mode entanglement can be significantly enhanced by increasing the particle number, provided the system is initialized within a specific symmetry sector of a strong symmetry operator. We identified the crucial role of these symmetry sectors in selecting physically relevant steady states from the extensively degenerate manifold permitted by strong symmetries and their associated conserved charges. In this sense, the strong symmetry structure not only governs the existence of steady states but also controls their entanglement properties, such as the persistence of the long-range correlations. Since the dephasing acts on the central site of an odd-sized chain, it preserves parity and confines the dynamics to the even-parity sector for an initially even state. The system therefore relaxes not to the full infinite-temperature state, but to the maximally mixed state within the even-parity subspace. As the parity eigenstates are coherent superpositions of mirror-related sites, this restricted identity is intrinsically nonlocal in the site basis and necessarily retains finite long-range correlations. Thus, the observed correlations are symmetry-protected remnants of the conserved parity, rather than coherences generated by the dissipative dynamics.

Our analytical results, supported by direct numerical simulations, provide a transparent framework for understanding how measurement-induced dephasing, when combined with symmetry constraints, can be harnessed to stabilize long-range entangled pairs. This highlights a broader paradigm in open quantum systems: rather than being purely detrimental, dissipation and measurements can be engineered to generate and protect nonlocal quantum resources.

In a nutshell, our results establish stochastic local measurements, when combined with symmetry constraints, as a powerful and conceptually minimal paradigm for engineering pairwise mode entanglement in fermionic lattices. This mechanism operates without the need for complex control protocols and instead leverages the intrinsic structure of measurement-induced dynamics and conserved quantities. In doing so, it provides a unifying perspective that bridges measurement-driven and dissipative approaches to quantum state engineering.

More broadly, our work identifies a fundamentally new route for generating robust bipartite entanglement over arbitrarily long distances using strictly local operations. In contrast to projective measurement-based schemes, such as that of \cite{Lu2026}, where a single measurement can directly project the system onto a target entangled state, our protocol generates entanglement continuously through monitoring of the
central site. Notably, no entangling gate is required; instead, the emergence of long-range entangled pairs results from the interplay between
local measurement-induced dynamics and the strong symmetries of the system. By elucidating the decisive role of strong symmetries and conserved charges in stabilizing nontrivial steady states, it deepens our understanding of structure and control in open quantum systems. These insights not only advance the theoretical foundations of nonequilibrium quantum dynamics but also open promising avenues for practical implementations in quantum technologies, including long-distance quantum communication, distributed quantum computation \cite{Lidar03, Kohout08}, and quantum-enhanced metrology \cite{Okane21, Kukita11}.

\appendix

\subsection*{Appendix A. Derivation of $\hat{\mathcal{Q}}=\hat{\nu}_{\mathrm{e}}-\hat{\nu}_{\mathrm{o}}$}
\label{Appendix_A}

We define the spatial reflection operation $\hat{\mathcal{R}}$ about the center of the lattice as
\begin{equation}
\hat{\mathcal{R}}|i\rangle = |N+1-i\rangle.
\end{equation}
Thus, the reflection operation maps a site on the left side of the lattice to the corresponding site on the right side and vice versa.

For a reflection-symmetric lattice Hamiltonian, the Hamiltonian
commutes with the reflection operator, $[\hat{H},\mathcal{\hat R}]=0$. Consequently, the single-particle eigenstates $|\psi_\alpha\rangle$ can be chosen to be simultaneous eigenstates of the reflection operator. Thus,
\begin{equation}
\mathcal{\hat R}|\psi_{\alpha}\rangle
=r_{\alpha}|\psi_{\alpha}\rangle,
\end{equation}
where the reflection eigenvalue satisfies
\begin{equation}
r_{\alpha}=
\begin{cases}
+1, & \text{for ${\alpha \in \mathrm{even}}$},\\
-1, & \text{for ${\alpha \in \mathrm{odd}}$}.
\end{cases}
\end{equation}

Writing the single-particle wavefunction in the site basis as: $\psi_{\alpha}(i)=\langle i|\psi_{\alpha}\rangle$, the reflection eigenvalue equation gives
\begin{equation}
\psi_{\alpha}(N+1-i)
=r_{\alpha}\psi_{\alpha}(i).
\label{eq:psi_prop}
\end{equation}

Next, we introduce the annihilation (creation) operator $\hat{a}_{\alpha}$ ($\hat{a}_{\alpha}^\dagger$) associated with
the single-particle eigenstate $|\psi_{\alpha}\rangle$. The site operators
can then be expanded as
\begin{equation}
\hat{f}_{i}
=
\sum_{\alpha}
\psi_{\alpha}(i)\hat{a}_{\alpha}
\quad
\text{and}
\quad
\hat{f}_{i}^{\dagger}
=
\sum_{\alpha}
\psi_{\alpha}^{*}(i)\hat{a}_{\alpha}^{\dagger}.
\end{equation}
Substituting these expressions into the definition of
$\hat{\mathcal{Q}}$, we obtain
\begin{equation}
\hat{\mathcal{Q}}=\sum_{i=1}^{N} \hat{f}_{i}^{\dagger} \hat{f}_{N+1-i}
=
\sum_{i=1}^{N}
\left[
\sum_{\alpha}
\psi_{\alpha}^{*}(i)
\hat{a}_{\alpha}^{\dagger}
\right]
\left[
\sum_{\beta}
\psi_{\beta}(N+1-i)
\hat{a}_{\beta}
\right].
\end{equation}
Expanding the products gives
\begin{equation}
\hat{\mathcal{Q}}
=
\sum_{\alpha,\beta}
\hat{a}_{\alpha}^{\dagger}
\hat{a}_{\beta}
\sum_{i=1}^{N}
\psi_{\alpha}^{*}(i)
\psi_{\beta}(N+1-i).
\end{equation}
Using Eq.~\eqref{eq:psi_prop}, we get
\begin{equation}
\hat{\mathcal{Q}}
=
\sum_{\alpha,\beta}
r_{\beta}
\hat{a}_{\alpha}^{\dagger}
\hat{a}_{\beta}
\sum_{i=1}^{N}
\psi_{\alpha}^{*}(i)
\psi_{\beta}(i)=\sum_{\alpha}
r_{\alpha}
\hat{a}_{\alpha}^{\dagger}
\hat{a}_{\alpha}
\end{equation}
where, in the last equality, we have used the orthonormality of the single-particle eigenstates. Now defining
\begin{equation}
    \hat\nu_e=\sum_{\alpha \in \mathrm{even}}\hat a_\alpha^\dagger \hat a_\alpha,
    \quad
    \hat\nu_o=\sum_{\alpha \in \mathrm{odd}}\hat a_\alpha^\dagger \hat a_\alpha
\end{equation}
and since ($r_{\alpha}=+1$) for even-parity states and ($r_{\alpha}=-1$) for odd-parity states, we get the expression for $\mathcal{\hat Q}$ in Eq.~\eqref{eq:Q_new}.

\subsection*{Appendix B. Uniqueness of the Single-Particle Steady State}

\label{Appendix_B}

For the single-particle sector, the lattice Hilbert space is
\begin{equation}
\mathcal H = \mathrm{span}\{|i\rangle\}_{i=1}^{N},
\end{equation}
where $|i\rangle = \hat f_i^\dagger |0\rangle$, and \(|0\rangle\) denotes the vacuum state. The even-parity subspace is
\begin{equation}
\mathcal H_+ =
\mathrm{span}\left\{
|1_+\rangle, |2_+\rangle,\dots, |(c-1)_+\rangle, |c\rangle
\right\},
\end{equation}
with
\begin{equation}
|j_+\rangle
=
\frac{|j\rangle + |N+1-j\rangle}{\sqrt2}.
\end{equation}

Since $\hat L_c = \hat n_c = |c\rangle\langle c|$, the jump operator acts entirely within \(\mathcal H_+\). Consequently, it is sufficient to restrict the analysis to the even-parity sector, consistent with the fact that odd-parity states remain unaffected by the central-site dephasing.

The condition $[\hat A,\hat L_c]=0$ implies that
\begin{equation}
\hat A|c\rangle = \eta |c\rangle ,
\label{eq:A_c}
\end{equation}
for some \(\eta\in\mathbb C\), and furthermore that \(\hat A\) cannot couple \(|c\rangle\) to any orthogonal state.

The Hamiltonian projected onto the even-parity sector takes the form
\begin{equation}
\hat H_+
=
-J\left(
\sum_{j=1}^{c-2}
|j_+\rangle\langle (j+1)_+|
+
\sqrt2\, |(c-1)_+\rangle\langle c|
\right)
+\mathrm{h.c.}
\end{equation}

Now imposing
\begin{equation}
[\hat A,\hat H_+]\,|c\rangle =0
\end{equation}
gives
\begin{equation}
\hat A\hat H_+|c\rangle
=
\hat H_+\hat A|c\rangle .
\end{equation}
Using Eq.~\eqref{eq:A_c}, one obtains
\begin{equation}
-\sqrt2 J\,\hat A |(c-1)_+\rangle
=
-\sqrt2 J\,\eta |(c-1)_+\rangle ,
\end{equation}
which immediately yields
\begin{equation}
\hat A |(c-1)_+\rangle
=
\eta |(c-1)_+\rangle .
\end{equation}

Next, imposing
\begin{equation}
[\hat A,\hat H_+]\,|(c-1)_+\rangle =0
\end{equation}
leads to
\begin{equation}
-J\,\hat A
\left(
\sqrt2\,|c\rangle + |(c-2)_+\rangle
\right)
=
-J\,\eta
\left(
\sqrt2\,|c\rangle + |(c-2)_+\rangle
\right),
\end{equation}
from which it follows that
\begin{equation}
\hat A |(c-2)_+\rangle
=
\eta |(c-2)_+\rangle .
\end{equation}

Iterating this argument recursively for all remaining basis states yields
\begin{equation}
\hat A |j_+\rangle
=
\eta |j_+\rangle,
\qquad
1\le j \le c-1.
\label{eq:A_recursive}
\end{equation}

Combining Eqs.~\eqref{eq:A_c} and \eqref{eq:A_recursive}, we conclude that
\begin{equation}
\hat A = \eta\,\hat{\mathbb I}_+,
\end{equation}
where \(\hat{\mathbb I}_+\) denotes the identity operator on the even-parity sector \(\mathcal H_+\). Therefore, the commutant of \(\{\hat H_+,\hat L_c\}\) within the even-parity subspace is trivial, implying that the steady state in this sector is unique.

\subsection*{Appendix C. Exact single-particle dynamical solution for $N=3$}

\label{Appendix_C}

The essential features of the dynamics can be understood from the simplest nontrivial case of $N=3$ under engineered dephasing. For ease of calculation, we will go to the energy basis. The Lindblad jump operator in the energy basis can be written as
\begin{equation}
\hat{L}_c
=|c\rangle\langle c| = \sum_{\alpha}|\psi_\alpha\rangle\langle\psi_\alpha|\,(|c\rangle\langle c|)\sum_{\beta}|\psi_\beta\rangle\langle\psi_\beta|
=\sum_{\alpha\beta}
\phi_{\alpha} \phi_{\beta}
|\psi_\alpha\rangle\langle\psi_\beta|.
\end{equation}

Substituting the above expression into Eq.~\eqref{eq:Dyn} and evaluating each contribution explicitly yields
\begin{equation}
\dot{\rho}_{\alpha\beta}
=
-i(\varepsilon_\alpha-\varepsilon_\beta)\rho_{\alpha\beta}
+2\gamma
\left(\phi_{\alpha} \phi_{\beta} \rho^{(l)}_{cc} - \frac{1}{2}
\sum_\delta\big(\phi_{\alpha} \phi_{\delta} \rho_{\delta\beta} + \phi_{\delta} \phi_{\beta} \rho_{\alpha\delta}\big)\right),
\label{eq:Dyn_Enrg}
\end{equation}
where $\rho_{\alpha\beta}=\langle\psi_\alpha|\hat\rho|\psi_\beta\rangle$ and $\rho^{(l)}_{cc}=\langle c|\hat\rho|c\rangle
=\sum_{\alpha\beta} \phi_\alpha \phi_\beta \rho_{\alpha\beta}$ where the superscript $(l)$ emphasizes that the density matrix element is in the lattice basis.

From Eq.~\eqref{eq:Dyn_Enrg}, the density matrix elements in the energy basis evolve as
\begin{equation}
\begin{pmatrix}
\dot \rho_{00} \\
\dot \rho_{22} \\
\dot \rho_{02} \\
\dot \rho_{20} 
\end{pmatrix}=\begin{pmatrix}
-\frac{\gamma}{2} & \frac{\gamma}{2} & 0 & 0 \\
\frac{\gamma}{2} & -\frac{\gamma}{2} & 0 & 0 \\
0 & 0 & -\frac{\gamma}{2} + i\Delta & \frac{\gamma}{2} \\
0 & 0 & \frac{\gamma}{2} & -\frac{\gamma}{2} - i\Delta
\end{pmatrix}
\begin{pmatrix}
\rho_{00} \\
\rho_{22} \\
\rho_{02} \\
\rho_{20} 
\end{pmatrix}
\end{equation}
where $\Delta = \varepsilon_2 - \varepsilon_0=2\sqrt{2}J$ is the energy gap between the ground and 2nd excited state. The dynamical matrix has two independent $2\times2$ blocks with eigenvalues
\begin{equation}
\theta_1 = 0, \quad 
\theta_2 = -\gamma, \quad
\theta_{3,4} = -\frac{\gamma}{2} \pm i\zeta,
\end{equation}
where
\begin{equation}
\zeta = \frac{1}{2}\sqrt{4\Delta^2 - \gamma^2}.
\end{equation}

The eigenvectors corresponding to the eigenvalues of the dynamical matrix are
\begin{align}
\vec{v}_1 &= \frac{1}{\sqrt{2}} (1,\,1,\,0,\,0)^T, \quad
\vec{v}_2 = \frac{1}{\sqrt{2}} (1,\,-1,\,0,\,0)^T, \\
\vec{v}_3 &= \frac{1}{n_1} (0,\,0,\,1,\,x)^T, \quad
\vec{v}_4 = \frac{1}{n_2} (0,\,0,\,1,\,y)^T.
\end{align}
where
\begin{equation}
x = -\frac{2i}{\gamma}(\Delta - \zeta), \quad y = -\frac{2i}{\gamma}(\Delta + \zeta)
\end{equation}
\begin{equation}
n_1 = \sqrt {1 + |x|^2}, \quad
n_2 = \sqrt {1 + |y|^2}.
\end{equation}

We construct the transformation matrix using the above eigenvectors as
\begin{equation}
S =
\begin{pmatrix}
\frac{1}{\sqrt{2}} & \frac{1}{\sqrt{2}} & 0 & 0 \\
\frac{1}{\sqrt{2}} & -\frac{1}{\sqrt{2}} & 0 & 0 \\
0 & 0 & \frac{1}{n_1} & \frac{1}{n_2} \\
0 & 0 & \frac{x}{n_1} & \frac{y}{n_2}
\end{pmatrix}
\end{equation}
which is invertible and has an inverse
\begin{equation}
S^{-1} =
\begin{pmatrix}
\frac{1}{\sqrt{2}} & \frac{1}{\sqrt{2}} & 0 & 0 \\
\frac{1}{\sqrt{2}} & -\frac{1}{\sqrt{2}} & 0 & 0 \\
0 & 0 & \frac{y n_1}{y-x} & -\frac{n_1}{y-x} \\
0 & 0 & -\frac{x n_2}{y-x} & \frac{n_2}{y-x}
\end{pmatrix}.
\end{equation}

Using this basis transformation, we can write
\begin{equation}
S^{-1}
\begin{pmatrix}
\dot{\rho}_{00} \\
\dot{\rho}_{22} \\
\dot{\rho}_{02} \\
\dot{\rho}_{20}
\end{pmatrix}
=
\begin{pmatrix}
0 & 0 & 0 & 0 \\
0 & -\gamma & 0 & 0 \\
0 & 0 & -\frac{\gamma}{2} + i\zeta & 0 \\
0 & 0 & 0 & -\frac{\gamma}{2} - i\zeta
\end{pmatrix}
S^{-1}
\begin{pmatrix}
\rho_{00} \\
\rho_{22} \\
\rho_{02} \\
\rho_{20}
\end{pmatrix}
\end{equation}

From this, we get the solution:
\begin{align}
\rho_{00}(t)
&= \frac{1}{2} \left[
1 + e^{-\gamma t} \rho_{00}(0)
- e^{-\gamma t} \rho_{22}(0)
\right] \\
\rho_{22}(t)
&= \frac{1}{2} \left[
1 - e^{-\gamma t} \rho_{00}(0)
+ e^{-\gamma t} \rho_{22}(0)
\right] \\
\rho_{02}(t) &= e^{-\gamma t/2}
\left[
\rho_{02}(0)\left(\cos \zeta t + i\frac{\Delta}{\zeta}\sin \zeta t\right)
+ \rho_{20}(0)\frac{\gamma}{2\zeta}\sin \zeta t
\right]= \rho_{20}^*(t),
\end{align}

We now consider the initial state to be a Fock state $|\chi(0)\rangle = |010\rangle$ which is symmetric with respect to the central site. This Fock state can be written in the energy basis as
\begin{equation}
|\Psi(0)\rangle
= \frac{1}{\sqrt{2}}\left(|\psi_0\rangle - |\psi_2\rangle\right).
\end{equation}
So we get,
\begin{equation}
\rho_{00}(0)=\rho_{22}(0)=\frac{1}{2}, \quad
\rho_{02}(0)=\rho_{20}(0)=-\frac{1}{2}.
\end{equation}

Substituting into the general solution, we obtain
\begin{equation}
\rho_{00}(t) = \rho_{22}(t) = \frac{1}{2},
\end{equation}
\begin{equation}
\rho_{02}(t) = -\frac{1}{2} e^{-\gamma t/2}
\left[
\cos(\zeta t)
+ i\frac{\Delta}{\zeta}\sin(\zeta t)
+ \frac{\gamma}{2\zeta}\sin(\zeta t)
\right]= \rho_{20}^*(t).
\end{equation}

The density matrix elements in the lattice site basis follow from
\begin{equation}
\rho_{ij}^{(l)}(t) = \langle i | \hat{\rho}(t) | j \rangle
= \sum_{\alpha,\beta}
\rho_{\alpha\beta}(t)\psi_\alpha(i)\psi_\beta(j)
\end{equation}
where, for this case, we have
\begin{equation}
    \psi_{0}(j)=\frac{1}{\sqrt2}\sin\left(\frac{\pi j}{4}\right),
\end{equation}
\begin{equation}
    \psi_{2}(j)=\frac{1}{\sqrt2}\sin\left(\frac{3 \pi j}{4}\right).
\end{equation}

Evaluating explicitly,
\begin{equation}
\rho_{11}^{(l)}(t)
= \frac{1}{4}
\left[
1 - e^{-\gamma t/2}
\left(\cos \zeta t + \frac{\gamma}{2\zeta}\sin \zeta t\right)\right],
\end{equation}

\begin{equation}
\rho_{13}^{(l)}(t)
= \rho_{31}^{(l)}(t)
= \rho_{33}^{(l)}(t)
= \rho_{11}^{(l)}(t),
\end{equation}

\begin{equation}
\rho_{22}^{(l)}(t)
= \frac{1}{2}
\left[
1 + e^{-\gamma t/2}
\left(\cos \zeta t + \frac{\gamma}{2\zeta}\sin \zeta t\right)
\right],
\end{equation}

\begin{equation}
\rho_{12}^{(l)}(t)
= \rho_{32}^{(l)}(t)
= e^{-\gamma t/2}\frac{\Delta}{2\sqrt{2}}\frac{i\sin \zeta t}{\zeta},
\end{equation}
\begin{equation}
\rho_{21}^{(l)}(t)
= \rho_{23}^{(l)}(t)
= \left[\rho_{12}^{(l)}(t)\right]^*.
\end{equation}

{\color{black} For $\gamma<2\Delta$, the nonstationary contributions exhibit damped oscillations with decay rate $\gamma/2$. At $\gamma=2\Delta$, the dynamics becomes critically damped, while for $\gamma>2\Delta$ the oscillatory behavior is replaced by overdamped relaxation. The diagonalization above is valid away from $\gamma=2\Delta$, where $\zeta=0$ and the eigenvector matrix $S$ becomes singular; nevertheless, the time-domain solution remains well-defined in the limit $\zeta\to0$. In all cases, the nonstationary contributions decay at long times, and the system approaches a unique steady state.} The corresponding steady-state density matrix, $\hat{\rho}^{\text{ss}}$, assumes a form of `X'-state with nonzero matrix elements $[\rho^{(l)}_{11}]^{\text{ss}} = [\rho^{(l)}_{13}]^{\text{ss}} = [\rho^{(l)}_{22}]^{\text{ss}}/2 = [\rho^{(l)}_{31}]^{\text{ss}} = [\rho^{(l)}_{33}]^{\text{ss}} = 1/4$. Hence, although the nearest-neighbor sites are bereft of any correlations in the steady state, a finite amount of correlation is developed between the distant end parties via Liouvillian dynamics under study by initiating the system in a simple Fock basis.

\subsection*{Appendix D. Derivation of the Reduced Density Matrix $\hat{\rho}_{\mathrm{rd}}(i,j)$}

\label{Appendix_D}

For a particle-number-conserving fermionic many-body system with $\mathcal{N}_{\!f}$ fermions, the two-site reduced density matrix is obtained by tracing out all degrees of freedom except those associated with sites \(i\) and \(j\),
\begin{equation}
\hat{\rho}_{\mathrm{rd}}(i,j)
=
\mathrm{Tr}_{\overline{\{i,j\}}}
\left(\hat{\rho}\right).
\end{equation}
We choose the local basis as
\begin{equation}
\{|00\rangle_{ij},\ |01\rangle_{ij},\ |10\rangle_{ij},\ |11\rangle_{ij}\},
\end{equation}
where for example \(|10\rangle_{ij}=|1\rangle_i  \otimes |0\rangle_j\) and the same thing follows for the other basis states.

The diagonal matrix elements are the expectation values of the projection operators onto the four basis states,
\begin{align}
|00\rangle_{ij}\langle00|_{ij} \nonumber
&=
(1-\hat n_i)(1-\hat n_j),\\
|01\rangle_{ij}\langle01|_{ij} \nonumber
&=
(1-\hat n_i)\hat n_j,\\
|10\rangle_{ij}\langle10|_{ij} \nonumber
&=
\hat n_i(1-\hat n_j),\\
|11\rangle_{ij}\langle11|_{ij}
&=
\hat n_i\hat n_j.
\end{align}
Therefore,
\begin{align}
\left[\hat{\rho}_{\mathrm{rd}}(i,j)\right]_{11} \nonumber
&=
1-\langle\hat n_i\rangle_{\mathcal{N}_{\!f}}-\langle\hat n_j\rangle_{\mathcal{N}_{\!f}}
+\langle\hat n_i\hat n_j\rangle_{\mathcal{N}_{\!f}},
\\
\left[\hat{\rho}_{\mathrm{rd}}(i,j)\right]_{22} \nonumber
&=
\langle\hat n_j\rangle_{\mathcal{N}_{\!f}}
-
\langle\hat n_i\hat n_j\rangle_{\mathcal{N}_{\!f}},
\\
\left[\hat{\rho}_{\mathrm{rd}}(i,j)\right]_{33} \nonumber
&=
\langle\hat n_i\rangle_{\mathcal{N}_{\!f}}
-
\langle\hat n_i\hat n_j\rangle_{\mathcal{N}_{\!f}},
\\
\left[\hat{\rho}_{\mathrm{rd}}(i,j)\right]_{44}
&=
\langle\hat n_i\hat n_j\rangle_{\mathcal{N}_{\!f}},
\end{align}
where $\left[\hat{\rho}_{\mathrm{rd}}(i,j)\right]_{11}$ means the first diagonal element of the reduced density matrix and the others follow similarly.

The only non-vanishing off-diagonal matrix elements connect the one-particle states \(|01\rangle\) and \(|10\rangle\). The corresponding operators are
\begin{equation}
|01\rangle_{ij}\langle10|_{ij}
=
\hat f_j^\dagger\hat f_i,
\qquad
|10\rangle_{ij}\langle01|_{ij}
=
\hat f_i^\dagger\hat f_j,
\end{equation}
which give
\begin{equation}
\left[\hat{\rho}_{\mathrm{rd}}(i,j)\right]_{23}
=
\langle\hat f_j^\dagger\hat f_i\rangle_{\mathcal{N}_{\!f}},
\qquad
\left[\hat{\rho}_{\mathrm{rd}}(i,j)\right]_{32}
=
\langle\hat f_i^\dagger\hat f_j\rangle_{\mathcal{N}_{\!f}}.
\end{equation}

All other off-diagonal matrix elements vanish because they connect states with different local particle numbers. Consequently, the reduced density matrix takes the form
\begin{equation}
\hat{\rho}_{\mathrm{rd}}(i,j)=
\begin{bmatrix}
1-\langle\hat n_i\rangle-\langle\hat n_j\rangle+\langle\hat n_i\hat n_j\rangle
&0&0&0\\
0&
\langle\hat n_j\rangle-\langle\hat n_i\hat n_j\rangle
&
\langle\hat f_j^\dagger\hat f_i\rangle
&0\\
0&
\langle\hat f_i^\dagger\hat f_j\rangle
&
\langle\hat n_i\rangle-\langle\hat n_i\hat n_j\rangle
&0\\
0&0&0&
\langle\hat n_i\hat n_j\rangle
\end{bmatrix}_{\!\mathcal{N}_{\!f}}.
\end{equation}
where the subscript ${\mathcal{N}_{\!f}}$ outside the matrix is used to emphasize that the expectation values are evaluated in the Hilbert space of ${\mathcal{N}_{\!f}}$ fermions.

For a non-interacting fermionic system, the Hamiltonian is quadratic in the fermionic creation and annihilation operators. Hence, the many-body state is Gaussian and Wick's theorem allows the four-point correlation function to be expressed in terms of two-point correlators,
\begin{align}
\langle\hat n_i\hat n_j\rangle_{\mathcal{N}_{\!f}}
&=
\left\langle
\hat f_i^\dagger
\hat f_i
\hat f_j^\dagger
\hat f_j
\right\rangle_{\mathcal{N}_{\!f}}
\nonumber\\
&=
\langle\hat f_i^\dagger\hat f_i\rangle_{\mathcal{N}_{\!f}}
\langle\hat f_j^\dagger\hat f_j\rangle_{\mathcal{N}_{\!f}}
-
\langle\hat f_i^\dagger\hat f_j^\dagger\rangle_{\mathcal{N}_{\!f}}
\langle\hat f_i\hat f_j\rangle_{\mathcal{N}_{\!f}}
+
\langle\hat f_i^\dagger\hat f_j\rangle_{\mathcal{N}_{\!f}}
\langle\hat f_i\hat f_j^\dagger\rangle_{\mathcal{N}_{\!f}}.
\end{align}
Since the Hamiltonian conserves particle number,
\begin{equation}
\langle\hat f_i\hat f_j\rangle_{\mathcal{N}_{\!f}}
=
\langle\hat f_i^\dagger\hat f_j^\dagger\rangle_{\mathcal{N}_{\!f}}
=
0,
\end{equation}
and, using the anticommutation relation $\{\hat f_i,\hat f_j^\dagger\}=\delta_{ij},$ we obtain, for \(i\neq j\),
\begin{equation}
\langle\hat f_i\hat f_j^\dagger\rangle_{\mathcal{N}_{\!f}}
=
-\langle\hat f_j^\dagger\hat f_i\rangle_{\mathcal{N}_{\!f}}.
\end{equation}
Therefore,
\begin{align}
\langle\hat n_i\hat n_j\rangle_{\mathcal{N}_{\!f}}
&=
\langle\hat n_i\rangle_{\mathcal{N}_{\!f}}
\langle\hat n_j\rangle_{\mathcal{N}_{\!f}}
-
\langle\hat f_i^\dagger\hat f_j\rangle_{\mathcal{N}_{\!f}}
\langle\hat f_j^\dagger\hat f_i\rangle_{\mathcal{N}_{\!f}}
\nonumber\\
&=
\langle\hat n_i\rangle_{\mathcal{N}_{\!f}}
\langle\hat n_j\rangle_{\mathcal{N}_{\!f}}
-
\left|
\langle\hat f_i^\dagger\hat f_j\rangle_{\mathcal{N}_{\!f}}
\right|^2,
\end{align}
where we have used $\langle\hat f_j^\dagger\hat f_i\rangle_{\mathcal{N}_{\!f}}
=
\langle\hat f_i^\dagger\hat f_j\rangle_{\mathcal{N}_{\!f}}^{*}$.

Thus, every element of the two-site reduced density matrix is completely determined by the one-body correlation functions $\langle\hat f_i^\dagger\hat f_j\rangle_{\mathcal{N}_{\!f}},
\langle\hat n_i\rangle_{\mathcal{N}_{\!f}},
\langle\hat n_j\rangle_{\mathcal{N}_{\!f}}.$

\end{document}